\documentclass[conference]{IEEEtran}
\usepackage{hyperref}

\ifCLASSINFOpdf
\else
\fi
\usepackage{amsmath}
\usepackage{amssymb}
\usepackage{amsmath}
\usepackage{xfrac}

\usepackage{tikz}
\usetikzlibrary{shapes}
\usetikzlibrary{matrix}
\usetikzlibrary{arrows}
\usetikzlibrary{decorations.markings}
\usetikzlibrary{spy}
\usetikzlibrary{backgrounds}
\usetikzlibrary{positioning}
\usetikzlibrary{calc}
\usepackage{pgfplots}
\usepgfplotslibrary{units}

\pgfplotsset{
    legend image with text/.style={
        legend image code/.code={%
            \node[anchor=center] at (0.3cm,0cm) {#1};
        }
    },
}

\definecolor{Set1-4-1}{RGB}{228,26,28}
\definecolor{Set1-4-2}{RGB}{55,126,184}
\definecolor{Set1-4-3}{RGB}{77,175,74}
\definecolor{Set1-4-4}{RGB}{152,78,163}
\definecolor{Set1-5-5}{RGB}{255,127,0}
\definecolor{Set1-7-1}{RGB}{228,26,28}
\definecolor{Set1-7-2}{RGB}{55,126,184}
\definecolor{Set1-7-3}{RGB}{77,175,74}
\definecolor{Set1-7-4}{RGB}{152,78,163}
\definecolor{Set1-7-5}{RGB}{255,127,0}
\definecolor{Set1-7-6}{RGB}{166,86,40}
\definecolor{Set1-7-7}{RGB}{0,0,0}

\newcommand{\figurewidth}{0.88}	% Width of (most) plots as a fraction of the column width
\newcommand{\figureheight}{0.75}% Height of (most) plots as a fraction of the column width

\IEEEoverridecommandlockouts
\begin{document}
\title{Polar Coding for the Large Hadron Collider: Challenges in Code Concatenation}

%\author{\IEEEauthorblockN{Alexios Balatsoukas-Stimming}\\
%Track: A. Communications Systems,\\
%Topic: 2. Coding and Decoding
%\IEEEauthorblockA{School of Electrical and\\Computer Engineering\\
%Georgia Institute of Technology\\
%Atlanta, Georgia 30332--0250\\
%Email: http://www.michaelshell.org/contact.html}
%}


\author{\IEEEauthorblockN{Alexios Balatsoukas-Stimming, %
\thanks{The authors thank the Hasler Foundation (\url{www.haslerstiftung.ch}) for financially supporting the presentation of this work at the 2017 Asilomar Conference on Signals, Systems, and Computers.}%
Tomasz Podzorny, %
Jan Uythoven}%
%\IEEEauthorblockA{\{alexios.balatsoukas, tomasz.podzorny, jan.uythoven\}@cern.ch}%
\IEEEauthorblockA{European Laboratory for Particle Physics (CERN), Geneva, Switzerland}%
}

%Telephone: (800) 555--1212, Fax: (888) 555--1212}

%\IEEEspecialpapernotice{Review Topic: A-2. Coding and Decoding}

\maketitle

\begin{abstract}
%We study a channel coding scheme for possible use in the beam interlock system of the Large Hadron Collider at CERN. This coding scheme uses a polar code concatenated with a long repetition code in order to protect the most critical bit within the codeword, while still providing good error protection for the remaining information bits. We explain why the most straightforward way to decode this concatenated coding scheme is ineffective and, using our observations, we propose an improved scheme that has a significant SNR gain in the examined scenarios.
In this work, we present a concatenated repetition-polar coding scheme that is aimed at applications requiring highly unbalanced unequal bit-error protection, such as the Beam Interlock System of the Large Hadron Collider at CERN. Even though this concatenation scheme is simple, it reveals significant challenges that may be encountered when designing a concatenated scheme that uses a polar code as an inner code, such as error correlation and unusual decision log-likelihood ratio distributions. We explain and analyze these challenges and we propose two ways to overcome them.
\end{abstract}


\section{Introduction}
The Large Hadron Collider (LHC) at the European Organization for Nuclear Research (CERN) collides two counter-rotating proton beams with a total energy in the order of $400$~MJ per proton beam. The combined beam energy is approximately equal to the kinetic energy of a Boeing 747-8 landing at its maximum landing weight ($312,072$ kg) with a typical landing speed of 135 knots ($70$ m/s). 

In case of a failure in any critical part of the LHC, the stored beam energy has to be released (i.e. ``dumped'') in a timely and controlled fashion in order to avoid damaging the accelerator. This is achieved by the Beam Interlock System (BIS)~\cite{Todd2006,Schmidt2006}, which broadcasts a single bit (called the ``beam permit''), summarizing the information from various monitors, to a specialized beam dumping system. A new version of the BIS is currently under study and one of the possibilities that are being explored is to make the system more flexible by transmitting additional bits of information over the link that is used to transmit the beam permit, mainly for low-latency monitoring purposes. The error rate requirements for the various bits are very different; the beam permit bit must be highly protected against errors in order to avoid false beam dumps and, more crucially, missing a dump request, while the integrity of the monitoring data is less crucial. This kind of unequal error protection can be achieved by using specialized error-correcting codes.

Polar codes~\cite{Arikan2009} are provably capacity-achieving channel codes with low complexity decoding algorithms. They are particularly attractive for our application as they do not exhibit error floors and they can be decoded very efficiently using successive cancellation (SC) decoding~\cite{Arikan2009}. Unfortunately, at practically interesting blocklengths the error-correcting performance of polar codes is often inferior to that of other modern channel coding schemes, such as low-density parity-check (LDPC) and Turbo codes. Attempts have been made to improve the error-correcting performance of polar codes by following various directions, including modified code construction, modified decoding algorithms, as well as concatenation of polar codes with other error-correcting codes. 

In the direction of concatenated codes, concatenations with both classical and modern error-correcting codes have been proposed. For example, the work of \cite{Seidl2010} considers the concatenation of polar codes with very short repetition and Hamming codes, \cite{Bakshi2010} examines the concatenation of polar codes with Reed-Solomon codes, and \cite{Wang2014} considers the concatenation of polar codes with Bose--Chaudhuri--Hocquenghem and convolutional codes. Finally, in the works of~\cite{Tal2015,Afisiadis2014}, polar codes are concatenated with a cyclic redundancy check code in the context of modified decoding algorithms. On the other hand, \cite{Guo2014,Zhang2014,Meng2017} use a concatenation of polar codes with LDPC codes, while \cite{Ye2016} presents a concatenation of polar codes with repeat-accumulate codes.

%When serially concatenating any code with a polar code, two interesting problems arise depending on whether the polar code is used as an outer code or as an inner (i.e., channel-facing) code. When the polar code is used as an outer code, it is unclear how the polar code should be designed as the channel the polar decoder is exposed to is a combination of the physical channel and the decoder for the inner code. When the polar code is used as an inner code, the code construction problem is shifted to the outer code.

%\subsubsection*{Contribution} In this work, we present a concatenated polar coding scheme that uses repetition codes and is aimed at applications that require highly unbalanced unequal bit-error protection, such as the LHC BIS described above. Even though this concatenation scheme is simple, it reveals several practical difficulties that may be encountered when designing a concatenated scheme that uses a polar code as an inner code, such as error correlation and unusual decision log-likelihood ratio distributions. We explain and analyze these challenges and we propose ways to address them.

\section{Background}
In this section, we provide background on the LHC BIS, as well as on the construction and decoding of polar codes.

\subsection{The Large Hadron Collider Beam Interlock System}
\begin{figure}[t]
	\centering
	\includegraphics[scale=0.28]{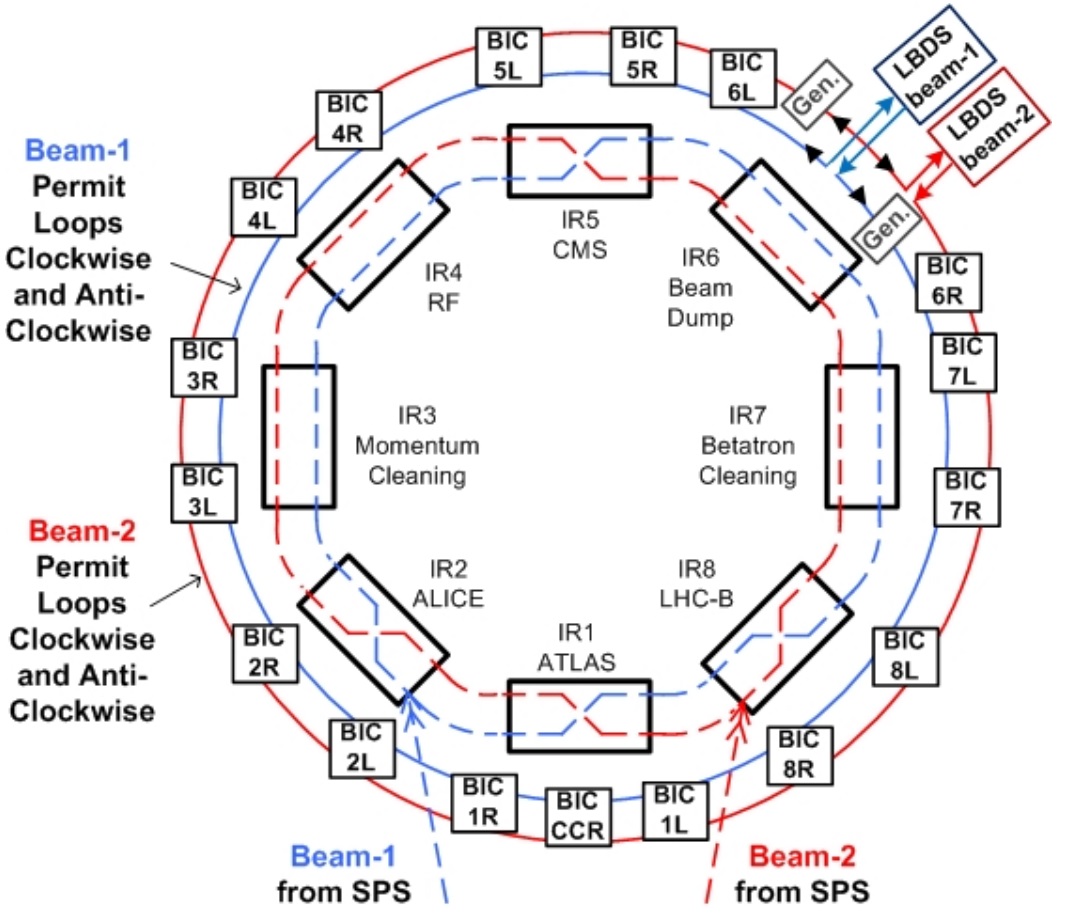}
	\caption{The Beam Interlock System of the Large Hadron Collider at CERN.}\label{fig:bis}
	\vspace{-0.1cm}
\end{figure}

As can be seen in Fig.~\ref{fig:bis}, the current BIS is a ring network that consists of $17$ nodes, called beam interlock controllers (BICs), that are spread around the $27$ kilometer circumference of the LHC and a generator node that generates and transmits a pre-defined frequency over dedicated optical fibers. In order to optimize the reliability and to minimize the propagation delay between any point of the ring network and the beam dumping system, there are in fact two distinct counter-directional optical ring networks. Each node passes this frequency on to its following node if and only if no operational faults have been detected and no beam dump has been requested by any of the equipment that is connected to the BIS. A loss of this frequency is detected by the beam dumping system, which then initiates the appropriate dumping procedure. In essence, the BIS uses digital frequency modulation in order to transmit a single bit of information, which is called the ``beam permit.''

\subsection{Polar Codes}
Polar codes are constructed by recursively applying a channel combining transformation to a channel $W$, followed by channel splitting~\cite{Arikan2009}. This results in $N = 2^n$ synthetic channels, denoted by $W_i(y_0^{N-1},u_0^{i-1}|u_i),~i=0,\hdots,{N-1}$. The synthetic channels have varying levels of reliability, which can be calculated using various methods~\cite{Arikan2009,Pedarsani2011,Tal2013}. A polar code of rate $R \triangleq \frac{K}{N},~0 < K < N,$ is obtained by letting the $K$ most reliable synthetic channels carry information bits, while freezing the input of the remaining channels to $0$. The set of non-frozen channel indices is denoted by $\mathcal{A}$ and the set of frozen channel indices is denoted by $\mathcal{A}^c$. The encoder generates a vector $u_0^{N-1}$ by setting $u_{\mathcal{A}^c}$ to $0$, while choosing $u_{\mathcal{A}}$ freely. A codeword is obtained as $x_0^{N-1} = u_{0}^{N-1}G_N,$ where $G_N$ is the generator matrix~\cite{Arikan2009}.

The SC decoding algorithm~\cite{Arikan2009} computes an estimate of $u_0$, denoted by $\hat{u}_0$, based on the received values $y_0^{N-1}$. Subsequently, $u_1$ is estimated using  $(y_0^{N-1},\hat{u}_0),$ etc. Let the log-likelihood ratio (LLR) for $W_i(y_0^{N-1},\hat{u}_0^{i-1}|u_i)$ be
\begin{align}
	L_i(y_0^{N-1},\hat{u}_0^{i-1}|u_i) \triangleq \log \left(\frac{W_i(y_0^{N-1},\hat{u}_0^{i-1}|u_i=0)}{W_i(y_0^{N-1},\hat{u}_0^{i-1}|u_i=1)}\right). 
\end{align}
In order to simplify notation, we will denote $L_i(y_0^{N-1},\hat{u}_0^{i-1}|u_i)$ by $L_i$ in the sequel. SC decoding decisions are taken according to
\begin{align}
	\hat{u}_i & =\left\{ \begin{matrix} 
							0, & L_i \geq 0 \text{ and } i \in \mathcal{A}, \\ 
							1, & L_i < 0 \text{ and } i \in \mathcal{A}, \\
							0, & i \in \mathcal{A}^c. \end{matrix} \right. \label{eq:scdec}
\end{align}
The decision LLRs $L_i$ can be calculated efficiently through a computation graph with complexity $O(N\log N)$~\cite{Arikan2009}. 

\section{Baseline Concatenated Repetition-Polar Scheme}
Consider the case where we want to transmit $K$ information bits per codeword and one of these bits is significantly more critical than the remaining bits. More specifically, let $b_{\text{crit}}$ denote the critical bit and let $b_{0}^{K-2}$ denote the remaining $K-1$ information bits. In the BIS of the LHC, $b_{\text{crit}}$ corresponds to the beam permit bit, while $b_{0}^{K-2}$ corresponds to other information (e.g., monitoring) that is less critical. In this section, we describe a baseline concatenated repetition-polar coding scheme that aims to solve the above problem and we evaluate its performance.

\subsection{Concatenated Repetition-Polar Scheme}
In order to improve the reliability of the critical bit $b_{\text{crit}}$, we use a concatenation scheme that first encodes $b_{\text{crit}}$ using a repetition code of length $k_{\text{rep}}$ and then encodes the resulting repetition codeword along with $b_{0}^{K-2}$ using a polar code, as shown in Fig.~\ref{fig:baselinescheme}. %This concatenation approach is similar to existing works~\cite{Seidl2010,Wang2014,Guo2014}, with the difference that existing works generally do not use repetition codes as outer codes.

\begin{figure}[t]
	\centering
	\begin{tikzpicture}[thick, scale=0.6, every node/.style={scale=0.9}]

	\small
	\tikzstyle{int}=[draw, fill=blue!15, minimum width=1.7cm, text width=1.7cm, minimum height=1.15cm, text centered]

	% Encoder
    \node [int] (rep) {Repetition Encoder};	
	\node [int] (polar) [below=0.75cm of rep] {Polar Encoder};
	\node (repin) [left=0.6cm of rep] {};
	\node (polarin) [below=1.57cm of repin] {};
    \draw[->] (repin) -- (rep) node [pos=0.25,label=above:{$b_{\text{crit}}$}] {};
	\draw[->] (polarin) -- (polar) node [pos=0.35,label=above:{$b_{0}^{K-2}$}] {};
	\draw[->] (rep.south) -- (polar) node [pos=0.5,label=right:{$u_{\mathcal{A}_{\text{crit}}}$}] {};

	% Channel
	\node [int] (chan) [right=1cm of polar] {Channel};
	\draw[->] (polar) -- (chan) node [pos=0.5,label=above:{$x_{0}^{N-1}$}] {};
	
	% Decoder
	\node [int] (polardec) [right=1cm of chan] {Polar Decoder};
	\node [int] (repdec) [above=0.75cm of polardec] {Soft Repetition Decoder};
	\node (repout) [right=0.6cm of repdec] {};
	\node (polarout) [right=0.6cm of polardec] {};
	\draw[->] (chan) -- (polardec) node [pos=0.5,label=above:{$y_{0}^{N-1}$}] {};
	\draw[->] (polardec) -- (repdec) node [pos=0.5,label=right:{$L_{\mathcal{A}_{\text{crit}}}$}] {};
	\draw[->] (repdec) -- (repout) node [pos=0.6,label=above:{$\hat{b}_{\text{crit}}$}] {};
	\draw[->] (polardec) -- (polarout) node [pos=0.75,label=above:{$\hat{b}_{0}^{K-2}$}] {};
	
\end{tikzpicture}
\vspace{-0.25cm}
	\caption{Baseliune concatenated repetition-polar coding scheme.}\label{fig:baselinescheme}
	\vspace{-0.1cm}
\end{figure}
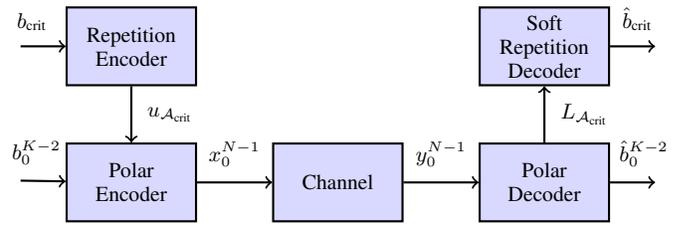

\begin{figure}[t]
	\centering
	\begin{tikzpicture}

	\pgfplotsset{grid style={solid}}

	\begin{semilogyaxis}[
		width = \figurewidth\columnwidth,
		height = \figureheight\columnwidth,
		xlabel = {\footnotesize $E_b/N_0$ (dB)},
		xlabel near ticks,
		ylabel = {\footnotesize Bit Error Rate},
		ylabel near ticks,
		xmin = 0, xmax = 6,
		ymin = 1e-6, ymax = 1e-1,
		tick label style={font=\footnotesize},
		grid = major,
		legend style={at={(.25,.015)},anchor=south,font=\scriptsize},
		legend cell align=left,
		legend columns=1,
	]		

		% k_rep = 1
		\addplot[Set1-4-1, opacity=0.75, ultra thick, dashdotted, mark=*, mark options={scale=0.7, solid, solid, opacity=1}] table {figures/data/AWGN_N128_SNR0.00_R0.50_SC_syst0_repLength1_repDecodingSoft_repType0_Quant0__Avg.dat};
		\addlegendentry{$\text{BER}_\text{avg} (k_{\text{rep}}=1$)}
		\addplot[Set1-4-1, opacity=0.75, ultra thick, mark=*, mark options={scale=0.7, solid, solid, opacity=1}] table {figures/data/AWGN_N128_SNR0.00_R0.50_SC_syst0_repLength1_repDecodingSoft_repType0_Quant0__Crit.dat};
		\addlegendentry{$\text{BER}_\text{crit} (k_{\text{rep}}=1$)}
		
		% k_rep = 5
		\addplot[Set1-4-2, opacity=0.75, ultra thick, dashdotted, mark=square*, mark options={scale=0.7, solid, solid, opacity=1}] table {figures/data/AWGN_N128_SNR0.00_R0.50_SC_syst0_repLength5_repDecodingSoft_repType0_Quant0__Avg.dat};
		\addlegendentry{$\text{BER}_\text{avg} (k_{\text{rep}}=5$)}
		\addplot[Set1-4-2, opacity=0.75, ultra thick, mark=square*, mark options={scale=0.7, solid, solid, opacity=1}] table {figures/data/AWGN_N128_SNR0.00_R0.50_SC_syst0_repLength5_repDecodingSoft_repType0_Quant0__Crit.dat};
		\addlegendentry{$\text{BER}_\text{crit} (k_{\text{rep}}=5$)}
		
		% k_rep = 11
		\addplot[Set1-4-3, opacity=0.75, ultra thick, dashdotted, mark=triangle*, mark options={scale=0.8, solid, solid, opacity=1}] table {figures/data/AWGN_N128_SNR0.00_R0.50_SC_syst0_repLength11_repDecodingSoft_repType0_Quant0__Avg.dat};
		\addlegendentry{$\text{BER}_\text{avg} (k_{\text{rep}}=11$)}		
		\addplot[Set1-4-3, opacity=0.75, ultra thick, mark=triangle*, mark options={scale=0.8, solid, solid, opacity=1}] table {figures/data/AWGN_N128_SNR0.00_R0.50_SC_syst0_repLength11_repDecodingSoft_repType0_Quant0__Crit.dat};
		\addlegendentry{$\text{BER}_\text{crit} (k_{\text{rep}}=11$)}
		
		\end{semilogyaxis}

\end{tikzpicture}%
\vspace{-0.25cm}
	\caption{Performance of an SC decoded polar code with blocklength $N=128$ and information rate $R_{\text{inf}}=\sfrac{1}{2}$ concatenated with a $k_{\text{rep}}$ soft-decision repetition code.}\label{fig:baseline}
	\vspace{-0.1cm}
\end{figure}
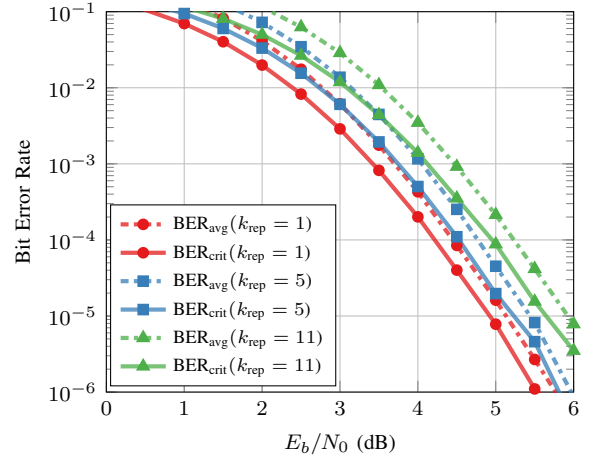

More specifically, we first construct a polar code of rate $R = \frac{K-1+k_{\text{rep}}}{N}$ and we denote the subset of $k_{\text{rep}}$ most reliable non-frozen channels by $\mathcal{A}_{\text{crit}} \subset \mathcal{A}$. Encoding is done by setting $u_{\mathcal{A}_{\text{crit}}} = b_{\text{crit}}$ and $u_{\mathcal{A}_{\text{non-crit}}} = b_{0}^{K-2}$, where $\mathcal{A}_{\text{non-crit}} = \mathcal{A} \backslash \mathcal{A}_{\text{crit}}$, and then using a polar encoder to compute $x_0^{N-1} = u_{0}^{N-1}G_N$. The effective information rate of this scheme is $R_{\text{inf}} = \frac{K}{N}$. The codeword $x_0^{N-1}$ is transmitted over the physical channel and a noisy version, denoted by $y_0^{N-1}$, is received.

At the receiver, we first decode the polar code using SC decoding and we then use the decision LLRs $L_{\mathcal{A}_{\text{crit}}}$ corresponding to the repetition codeword to decode the repetition code using standard soft decoding as
\begin{align}
	\hat{b}_{\text{crit}}	& = \left\{ \begin{matrix} 
											0, & \sum _{i \in \mathcal{A}_{\text{crit}}} L_i \geq 0, \\
											1, & \text{otherwise}.
										\end{matrix} \right. 
\end{align}

\subsection{Performance Evaluation}
In Fig.~\ref{fig:baseline} we present the performance of the concatenated coding scheme described in the previous subsection for a polar code with $N = 128$ and $R_{\text{inf}} = 0.5$ for various values of $k_{\text{rep}}$. We are interested in short polar codes due to the stringent latency requirements of the BIS. The simulations are performed for transmission over an AWGN channel using BPSK modulation and we denote the bit error rate (BER) of the critical bit $b_{\text{crit}}$ by $\text{BER}_\text{crit}$ and the average BER over both $b_{\text{crit}}$ and $b_{0}^{N-2}$ by $\text{BER}_\text{avg}$.

First, we observe that for $k_{\text{rep}} = 1$, $\text{BER}_\text{crit}$ is already slightly lower than $\text{BER}_\text{avg}$ since polar codes inherently have unequal error protection due to the varying reliabilities of the synthetic channels. We also observe that, as $k_{\text{rep}}$ increases, $\text{BER}_\text{avg}$ also increases, which is expected since, in order to keep the same effective information rate $R_{\text{inf}}$ for increasing $k_{\text{rep}}$, the rate $R$ of the underlying polar code has to be increased. However, we also observe that $\text{BER}_\text{crit}$ also increases with inreasing $k_{\text{rep}}$. The latter observation is counter-intuitive and unfortunately goes in the opposite direction of what we would like to achieve. However, as we will show in the following section, the observed behavior can be explained and counteracted.

\begin{figure}[t]
	\centering
	\includegraphics{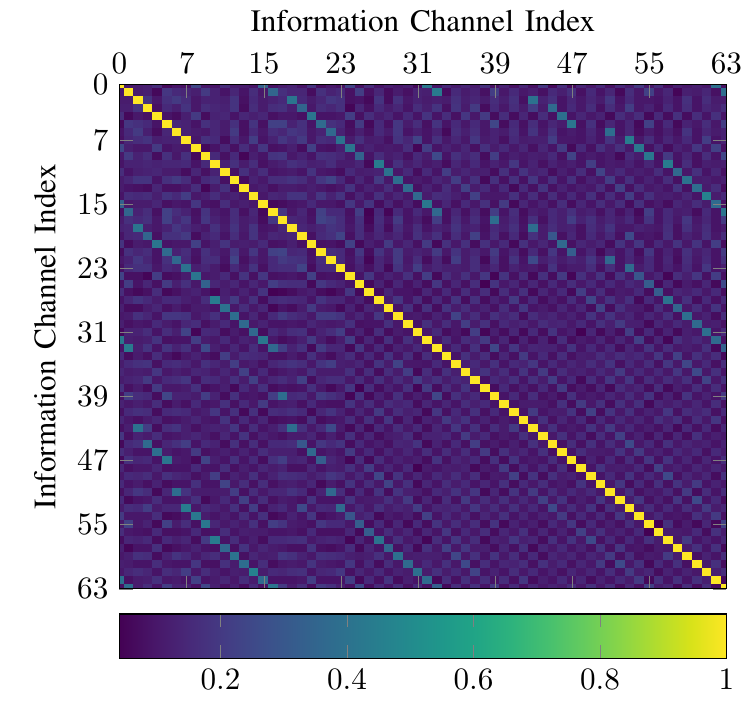}
	\caption{Correlation matrix of the error vector $\textbf{e}$ for a polar code with $N = 128$ and $R = 0.5$ at a design SNR of $0$~dB.}\label{fig:correlationbaseline}
	\vspace{-0.1cm}
\end{figure}

%\begin{figure}[t]
%	\centering
%	\includegraphics[scale=0.45]{figures/correlationbaseline.png}
%	\caption{Correlation matrix of the error vector $\textbf{e}$ for a polar code with $N = 128$ and $R_{\text{inf}} = 0.5$ at a design SNR of $0$~dB.}\label{fig:correlationbaseline}
%\end{figure}

\section{Issues in the Baseline Concatenated Scheme}
In this section, we identify two reasons why the repetition code is ineffective when concatenated with a polar code, namely the correlation between the errors in the decoded information bits in a polar code and the atypical distribution of the decision LLRs $L_i$.

\subsection{Correlation Between the Decoded Information Bits}
It has been shown that, in the case of the binary erasure channel, the errors in the information bits that are decoded by the SC decoder become uncorrelated as the blocklength goes to infinity~\cite{BastaniParizi2013}. However, at short-to-moderate blocklengths, the errors are generally highly correlated. This correlation reduces the diversity order of the repetition code and renders it less effective. 

In order to demonstrate this effect, let us define each element $e_i,~i\in\mathcal{A}$, of the error vector $\textbf{e}$ as
\begin{align}
	e_i	& = \left\{ \begin{matrix}
						0, & \hat{u}_i = u_i,\\
						1, & \text{otherwise}.
					\end{matrix}
			\right.
\end{align}
In Fig.~\ref{fig:correlationbaseline}, we show the correlation matrix of the error vector $\textbf{e}$ for a polar code with $N = 128$ and $R = 0.5$, which is obtained by running simulations at the design SNR of $0$~dB, which translates to an Eb/N0 of $3$~dB for $R=0.5$. We observe that there exist several large correlation coefficients, especially between error vector elements corresponding to synthetic channels that are  close in terms of their index $i$.

\begin{figure}[t]
	\centering
	\begin{tikzpicture}[scale=0.55]
\large
\begin{axis}[
    xmin = -350, xmax = 350,
	xlabel={$D_{120}$ values},
	xlabel near ticks,
    ymin=1e-5, ymax=1,
	ylabel={Probability},
	ylabel near ticks,
    area style,
	ymode = log, 
	log origin=infty	
    ]
	\addplot+[ybar interval] table {figures/data/histNonSyst120.dat};
\end{axis}%
\end{tikzpicture}%
\hspace{-0.55cm}%
\begin{tikzpicture}[scale=0.55]%
\large%
\begin{axis}[
    xmin = -350, xmax = 350,
	xlabel={$D_{128}$ values},
	xlabel near ticks,
    ymin=1e-5, ymax=1,
	yticklabels={\empty},
    area style,
	ymode = log, 
	log origin=infty
    ]
	\addplot+[ybar interval,fill=blue!50] table {figures/data/histNonSyst128.dat};
\end{axis}%
\end{tikzpicture}%
\vspace{-0.25cm}
	\caption{Histograms of $D_{120}$ (left) and $D_{128}$ (right) for a polar code with $N = 128$ and $R = 0.5$ at a design SNR of $0$~dB.}\label{fig:histNonSyst}
	\vspace{-0.1cm}
\end{figure}
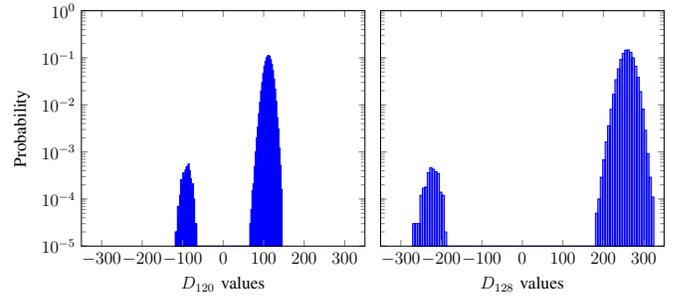

\subsection{Distribution of the Decision LLRs}\label{sec:llrhist}
Let us define $D_i$ as
\begin{align}
	D_i	& = \left\{ \begin{matrix}
						L_i, & u_i = 0,\\
						-L_i, & \text{otherwise},
					\end{matrix}
			\right.
\end{align}
so that $D_i \geq 0$ means that the decision for bit $i$ was correct (i.e., $e_i = 0$), and $D_i < 0$ means that the decision for bit $i$ was erroneous. 

In Fig.~\ref{fig:histNonSyst}, we show histograms of $D_{120}$ and $D_{128}$ for a polar code with $N = 128$ and $R_{\text{inf}} = 0.5$, which is obtained by running simulations at the design SNR of $0$~dB. Synthetic channel $i=128$ is the most reliable channel, while synthetic channel $i=120$ is the 5th most reliable channel, meaning that these two channels will belong to $\mathcal{A}_{\text{crit}}$ for both $k_{\text{rep}}=5$ and $k_{\text{rep}}=11$. We use a logarithmic vertical axis for presentation clarity. 

We observe that, in both cases, erroneous decisions are not caused by long tails in the distributions, but they actually come from erroneous LLR values that have very large magnitudes. These overconfident erroneous decisions can cause significant problems in the context of a soft repetition decoder. For example, an erroneous value of $L_{128} = 250$ when $u_{128} = 1$ (equivalently, $D_{128} = -250$) is very unlikely to be counteracted by any correct value of $L_{120}$, since from Fig.~\ref{fig:histNonSyst} we can see that correct values for $L_{120}$ are very tightly concentrated around a value of approximately $100$.

\begin{figure}[t]
	\centering
	\includegraphics{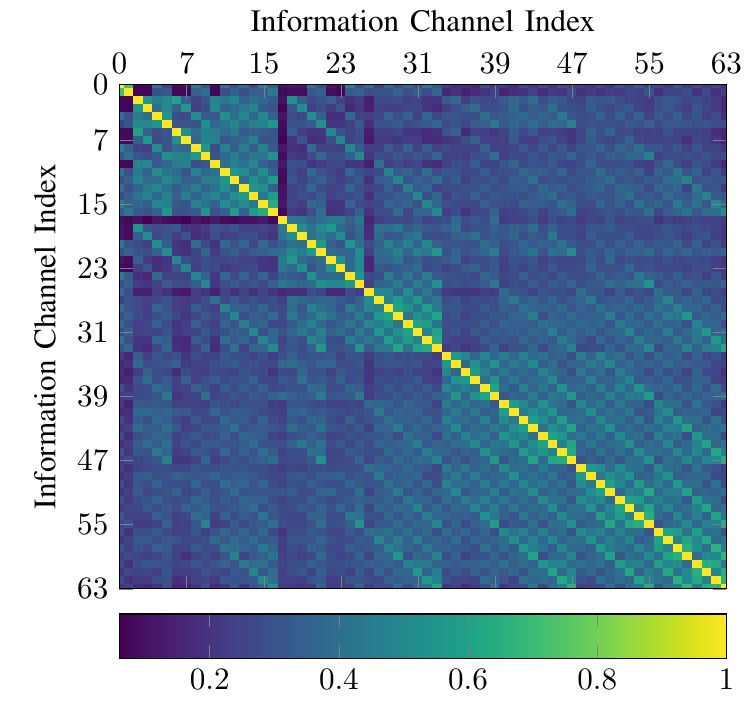}
	\caption{Correlation matrix of the error vector $\textbf{e}$ for a systematic polar code with $N = 128$ and $R = 0.5$  at a design SNR of $0$~dB.}\label{fig:correlationimproved}
	\vspace{-0.1cm}
\end{figure}

%\begin{figure}[t]
%	\centering
%	\includegraphics[scale=0.45]{figures/correlationimproved.png}
%	\caption{Correlation matrix of the error vector $\textbf{e}$ for a systematic polar code with $N = 128$ and $R_{\text{inf}} = 0.5$  at a design SNR of $0$~dB.}\label{fig:correlationimproved}
%\end{figure}

\section{Improved Concatenated Repetition-Polar Scheme}
In this section, we explain our proposed improvements to the baseline concatenated repetition-polar scheme that alleviate the two problems identified above. More specifically, we describe a scheme that uses systematic polar coding and hard repetition decoding code that significantly improves the error-correcting performance with respect to the baseline scheme.

\subsection{Error Vector Correlation} We note that the correlation problem has also been alluded to in other recent works~\cite{Wang2014,Meng2017}. However, the proposed solutions in these works rely on spreading the codeword bits for the outer code (in our case, the repetition code) over many codewords of the inner code (in our case, the polar code). This approach can make the codeword bits of the outer code completely uncorrelated, but it has a very high cost in terms of both the decoding latency and the memory required for codeword buffering. 

\begin{figure}[t]
	\centering
	\begin{tikzpicture}

	\pgfplotsset{grid style={solid}}

	\begin{semilogyaxis}[
		width = \figurewidth\columnwidth,
		height = \figureheight\columnwidth,
		xlabel = {\footnotesize Channel index within $\mathcal{A}$},
		xlabel near ticks,
		xtick={0,7,15,23,31,39,47,55,63},
		ylabel = {\footnotesize Bit Error Rate},
		ylabel near ticks,
		xmin = 0, xmax = 63,
		ymin = 1e-3, ymax = 1e-1,
		tick label style={font=\footnotesize},
		grid = major,
		legend style={at={(.3,.78)},anchor=south,font=\scriptsize},
		legend cell align=left,
		legend columns=1,
	]		

		% Per-channel BER
		\addplot[Set1-4-1, opacity=0.75, ultra thick, only marks, mark=*, mark options={scale=0.7, solid, solid, opacity=1}] table {figures/data/BERNonSyst.dat};
		\addlegendentry{Non-Systematic Polar Code}
		\addplot[Set1-4-2, opacity=0.75, ultra thick, only marks, mark=square*, mark options={scale=0.7, solid, solid, opacity=1}] table {figures/data/BERSyst.dat};
		\addlegendentry{Systematic Polar Code}
		
		\end{semilogyaxis}

\end{tikzpicture}%
\vspace{-0.25cm}
	\caption{Per-channel information bit BER for a non-systematic and a systematic polar code with blocklength $N=128$ and information rate $R_{\text{inf}}=\sfrac{1}{2}$.}\label{fig:perChannelBER}
	\vspace{-0.1cm}
\end{figure}
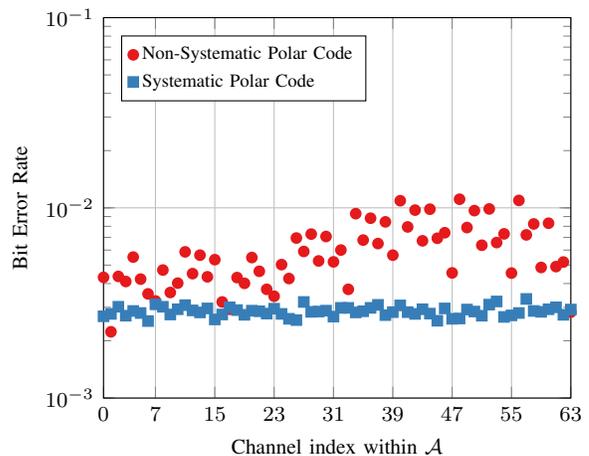

Our proposed solution, on the other hand, is to use systematic polar coding~\cite{Arikan2011}, which has a very small overhead in terms of both latency and complexity. More specifically, in our simulations, we have observed that systematic polar coding, apart from slightly improving the average BER, also significantly reduces the pair-wise correlations of the elements of the error vector~$\textbf{e}$. This effect can be clearly seen when comparing Fig.~\ref{fig:correlationbaseline} with Fig.~\ref{fig:correlationimproved}, where we show the correlation matrix of the error vector $\textbf{e}$ for a systematic polar code with $N = 128$ and $R_{\text{inf}} = 0.5$, which is obtained by running simulations at the design SNR of $0$~dB.

In Fig.~\ref{fig:improvedBER}, we observe that there is a significant gain of approximately $1.5$~dB for $\text{BER}_\text{crit}$ when moving from a non-systematic polar code to a systematic polar code while the gain for $\text{BER}_\text{avg}$ is only approximately $0.25$~dB, thus clearly demonstrating the decorrelating effect of systematic polar coding.

\subsection{LLR Distribution}
We have shown that systematic polar coding largely alleviates the error vector correlation problem. However, the LLR distribution problem remains even when systematic polar coding is used.\footnote{We note that the distributions of $D_i$ for systematic polar coding can be obtained by performing a soft encoding step on the SC decoder decision LLRs $L_i$. These distributions were not shown in Section~\ref{sec:llrhist} due to space limitations, but they are practically identical to the distributions of $D_i$ for non-systematic polar coding.} Moreover, as shown in Fig.~\ref{fig:perChannelBER}, when using systematic coding the BERs of the information channels are very similar. This means that the expected LLR magnitude $\mathbb{E}\left[|L_i|\right]$ for channel $i$ is not a good indicator for its reliability and soft repetition decoding is mismatched to the actual BER of each bit. 

%\begin{figure}[t]
%	\centering
%	\input{figures/improved_scheme.tikz}
%	\caption{Improved concatenated repetition-polar coding scheme.}\label{fig:improvedscheme}
%\end{figure}

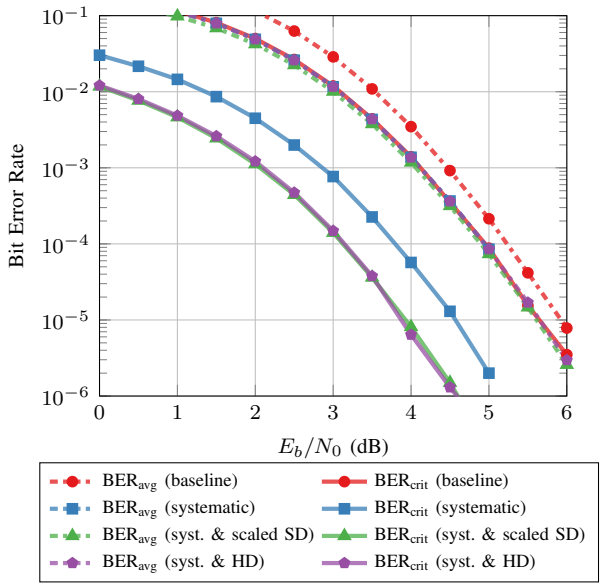
\begin{figure}[t]
	\centering
	\begin{tikzpicture}

	\pgfplotsset{grid style={solid}}

	\begin{semilogyaxis}[
		width = \figurewidth\columnwidth,
		height = \figureheight\columnwidth,
		xlabel = {\footnotesize $E_b/N_0$ (dB)},
		xlabel near ticks,
		ylabel = {\footnotesize Bit Error Rate},
		ylabel near ticks,
		xmin = 0, xmax = 6,
		ymin = 1e-6, ymax = 1e-1,
		tick label style={font=\footnotesize},
		grid = major,
		legend style={at={(.47,-0.49)},anchor=south,font=\scriptsize},
		legend cell align=left,
		legend columns=2,
	]		

		% k_rep 11, baseline
		\addplot[Set1-4-1, opacity=0.75, ultra thick, dashdotted, mark=*, mark options={scale=0.7, solid, solid, opacity=1}] table {figures/data/AWGN_N128_SNR0.00_R0.50_SC_syst0_repLength11_repDecodingSoft_repType0_Quant0__Avg.dat};
		\addlegendentry{$\text{BER}_\text{avg}$ (baseline)}
		\addplot[Set1-4-1, opacity=0.75, ultra thick, mark=*, mark options={scale=0.7, solid, solid, opacity=1}] table {figures/data/AWGN_N128_SNR0.00_R0.50_SC_syst0_repLength11_repDecodingSoft_repType0_Quant0__Crit.dat};
		\addlegendentry{$\text{BER}_\text{crit}$ (baseline)}
		
		% k_rep = 11, systematic
		\addplot[Set1-4-2, opacity=0.75, ultra thick, dashdotted, mark=square*, mark options={scale=0.7, solid, solid, opacity=1}] table {figures/data/AWGN_N128_SNR0.00_R0.50_SC_syst1_repLength11_repDecodingSoft_repType0_Quant0__Avg.dat};
		\addlegendentry{$\text{BER}_\text{avg}$ (systematic)}
		\addplot[Set1-4-2, opacity=0.75, ultra thick, mark=square*, mark options={scale=0.7, solid, solid, opacity=1}] table {figures/data/AWGN_N128_SNR0.00_R0.50_SC_syst1_repLength11_repDecodingSoft_repType0_Quant0__Crit.dat};
		\addlegendentry{$\text{BER}_\text{crit}$ (systematic)}		
		
		% k_rep = 11, systematic
		\addplot[Set1-4-3, opacity=0.75, ultra thick, dashdotted, mark=triangle*, mark options={scale=0.8, solid, solid, opacity=1}] table {figures/data/AWGN_N128_SNR0.00_R0.50_SC_syst1_repLength11_repDecodingSoft_repType0_Quant0_LLRscaling_Avg.dat};
		\addlegendentry{$\text{BER}_\text{avg}$ (syst. \& scaled SD)}
		\addplot[Set1-4-3, opacity=0.75, ultra thick, mark=triangle*, mark options={scale=0.8, solid, solid, opacity=1}] table {figures/data/AWGN_N128_SNR0.00_R0.50_SC_syst1_repLength11_repDecodingSoft_repType0_Quant0_LLRscaling_Crit.dat};
		\addlegendentry{$\text{BER}_\text{crit}$ (syst. \& scaled SD)}	
		
		% k_rep = 11, systematic and hard decoding
		\addplot[Set1-4-4, opacity=0.75, ultra thick, dashdotted, mark=pentagon*, mark options={scale=0.7, solid, solid, opacity=1}] table {figures/data/AWGN_N128_SNR0.00_R0.50_SC_syst1_repLength11_repDecodingHard_repType0_Quant0__Avg.dat};
		\addlegendentry{$\text{BER}_\text{avg}$ (syst. \& HD)}
		\addplot[Set1-4-4, opacity=0.75, ultra thick, mark=pentagon*, mark options={scale=0.7, solid, solid, opacity=1}] table {figures/data/AWGN_N128_SNR0.00_R0.50_SC_syst1_repLength11_repDecodingHard_repType0_Quant0__Crit.dat};
		\addlegendentry{$\text{BER}_\text{crit}$ (syst. \& HD)}		
		
		\end{semilogyaxis}

\end{tikzpicture}%
\vspace{-0.25cm}
	\caption{Performance of a systematic polar code with blocklength $N=128$ and information rate $R_{\text{inf}}=\sfrac{1}{2}$ concatenated with a $k_{\text{rep}} = 11$ repetition code.}\label{fig:improvedBER}
	\vspace{-0.1cm}
\end{figure}

One possible solution is to estimate $\mathbb{E}\left[|L_i|\right],~i \in \mathcal{A}$, at some SNR using simulations, and to scale the decision LLRs as $L_i/\mathbb{E}\left[|L_i|\right]$ so that all LLRs used by the soft repetition decoder are on a similar scale. As can be observed in Fig.~\ref{fig:improvedBER}, where we estimated $\mathbb{E}\left[|L_i|\right]$ at the polar code design SNR of $0$~dB, using soft decoding with LLR scaling in conjuction with systematic polar coding provides a further SNR gain of slightly more than $0.5$~dB with respect to the case where only systematic polar coding is used. 

However, as we demonstrated in Section~\ref{sec:llrhist}, erroneous bit decisions almost always stem from completely wrong LLR values and the distribution of the decision LLR magnitudes is highly concentrated. This means that the actual decision LLR magnitude carries very little information about the reliability of the corresponding bit decision, even when scaling the LLRs as explained previously. For this reason, as we can observe in Fig.~\ref{fig:improvedBER}, hard decision decoding has practically the same performance as soft decision decoding in this particular scenario, while being significantly less complex to implement. 

\subsection{Overall Performance Evaluation}
In Fig.~\ref{fig:systematicharddecision} we show the comparison that we showed for the baseline concatenated repetition-polar scheme in Fig.~\ref{fig:baseline}, but this time for the improved scheme. We observe that, with the improved scheme, we have successfully made $\text{BER}_\text{crit}$ decrease as $k_{\text{rep}}$ increases. At the same time,  the degradation of $\text{BER}_\text{avg}$ as $k_{\text{rep}}$ is increased is also smaller than in the baseline scheme.

\section{Conclusions}
In this work, we examined the concatenation of a repetition code with a polar code. More specifically, we showed that a baseline concatenated scheme that uses a non-systematic polar code and a soft repetition decoder performs particularly poorly. We identified two major reasons that lead to this poor performance, namely the error vector correlation and the unusual decision LLR distributions, and we proposed an improved scheme that employs two ways to overcome these problems, namely using systematic polar coding and hard repetition decoding. Our simulation results show the improved scheme has an SNR gain of up to $2$~dB for a $k_{\text{rep}} = 11$ repetition code concatenated with polar code with $N=128$ and $R_{\text{inf}} = 0.5$.

While the examined concatenated scheme is quite specific, the identified problems also affect other concatenated schemes that use a polar code as an inner code. In particular, systematic polar coding is an effective correlation-breaking approach that can be beneficial for any concatenated scheme. Moreover, because of the unusual decision LLR distributions that stem from the polarizing transformation that is used to construct the synthetic channels, low-complexity hard decision decoding may also be sufficient for other outer codes, especially as the blocklength is increased and the synthetic channels become more polarized.

\begin{figure}[t]
	\centering
	\begin{tikzpicture}

	\pgfplotsset{grid style={solid}}

	\begin{semilogyaxis}[
		width = \figurewidth\columnwidth,
		height = \figureheight\columnwidth,
		xlabel = {\footnotesize $E_b/N_0$ (dB)},
		xlabel near ticks,
		ylabel = {\footnotesize Bit Error Rate},
		ylabel near ticks,
		xmin = 0, xmax = 6,
		ymin = 1e-6, ymax = 1e-1,
		tick label style={font=\footnotesize},
		grid = major,
		legend style={at={(.5,-0.44)},anchor=south,font=\scriptsize},
		legend cell align=left,
		legend columns=2,
	]		

		% k_rep = 1
		\addplot[Set1-4-1, opacity=0.75, ultra thick, dashdotted, mark=*, mark options={scale=0.7, solid, solid, opacity=1}] table {figures/data/AWGN_N128_SNR0.00_R0.50_SC_syst1_repLength1_repDecodingHard_repType0_Quant0__Avg.dat};
		\addlegendentry{$\text{BER}_\text{avg} (k_{\text{rep}}=1$)}
		\addplot[Set1-4-1, opacity=0.75, ultra thick, mark=*, mark options={scale=0.7, solid, solid, opacity=1}] table {figures/data/AWGN_N128_SNR0.00_R0.50_SC_syst1_repLength1_repDecodingHard_repType0_Quant0__Crit.dat};
		\addlegendentry{$\text{BER}_\text{crit} (k_{\text{rep}}=1$)}
		
		% k_rep = 5		
		\addplot[Set1-4-2, opacity=0.75, ultra thick, dashdotted, mark=square*, mark options={scale=0.7, solid, solid, opacity=1}] table {figures/data/AWGN_N128_SNR0.00_R0.50_SC_syst1_repLength5_repDecodingHard_repType0_Quant0__Avg.dat};
		\addlegendentry{$\text{BER}_\text{avg} (k_{\text{rep}}=5$)}
		\addplot[Set1-4-2, opacity=0.75, ultra thick, mark=square*, mark options={scale=0.7, solid, solid, opacity=1}] table {figures/data/AWGN_N128_SNR0.00_R0.50_SC_syst1_repLength5_repDecodingHard_repType0_Quant0__Crit.dat};
		\addlegendentry{$\text{BER}_\text{crit} (k_{\text{rep}}=5$)}

		% k_rep = 11		
		\addplot[Set1-4-3, opacity=0.75, ultra thick, dashdotted, mark=triangle*, mark options={scale=0.8, solid, solid, opacity=1}] table {figures/data/AWGN_N128_SNR0.00_R0.50_SC_syst1_repLength11_repDecodingHard_repType0_Quant0__Avg.dat};
		\addlegendentry{$\text{BER}_\text{avg} (k_{\text{rep}}=11$)}
		\addplot[Set1-4-3, opacity=0.75, ultra thick, mark=triangle*, mark options={scale=0.8, solid, solid, opacity=1}] table {figures/data/AWGN_N128_SNR0.00_R0.50_SC_syst1_repLength11_repDecodingHard_repType0_Quant0__Crit.dat};
		\addlegendentry{$\text{BER}_\text{crit} (k_{\text{rep}}=11$)}		
		
		\end{semilogyaxis}

\end{tikzpicture}%
\vspace{-0.25cm}
	\caption{Performance of a systematic polar code with blocklength $N=128$ and information rate $R_{\text{inf}}=\sfrac{1}{2}$ concatenated with a $k_{\text{rep}}$ hard-decision repettition code.}\label{fig:systematicharddecision}
\end{figure}

\bibliographystyle{IEEEtran}
\bibliography{IEEEabrv,bibliography}

\end{document}